\documentclass[prl,twocolumn,superscriptaddress,showpacs,preprintnumbers,amsmath,amssymb,a4paper]{revtex4}
\usepackage{amssymb}

\usepackage[top=0.85in,bottom=0.9in,,left=0.75in,right=0.75in]{geometry}
\usepackage{graphicx}
\usepackage{dcolumn}
\usepackage{bm}
\usepackage{epsfig}
\newcommand{\ignore}[1]{}

\begin{document}

\title{Tuning Transport Properties of Topological Edge States of Bi(111) Bilayer Film by Edge Adsorption}

\author{Z. F. Wang}
\affiliation{Department of Materials Science and Engineering, University of Utah, Salt Lake City, UT 84112}

\author{Li Chen}
\thanks{E-mail: lchen.lyu@gmail.com}
\affiliation{Institute of Condensed Matter Physics, Linyi University, Linyi, Shandong 276005, China}
\affiliation{State Key Laboratory of Low Dimensional Quantum Physics and Department of Physics, Tsinghua University, Beijing 100084, China}

\author{Feng Liu}
\thanks{E-mail: fliu@eng.utah.edu}
\affiliation{Department of Materials Science and Engineering, University of Utah, Salt Lake City, UT 84112}
\affiliation{Collaborative Innovation Center of Quantum Matter, Beijing, China}

\begin{abstract}
Based on first-principles and tight-binding calculations, we report that the transport properties of
topological edge states of zigzag Bi(111) nanoribbon can be significantly tuned by H edge adsorption.
The Fermi velocity is increased by one order of magnitude, as the Dirac point is moved from Brillouin
zone boundary to Brillouin zone center and the real-space distribution of Dirac states are made twice more
delocalized. These intriguing changes are explained by an orbital filtering effect of edge H atoms,
which removes certain components of $p$ orbits of edge Bi atoms that reshapes the
topological edge states. In addition, the spin texture of the Dirac states is also modified, which
is described by introducing an effective Hamiltonian. Our findings not only are of fundamental interest
but also have practical implications in potential applications of topological insulators.

\end{abstract}


\maketitle
Studies of chemical and structural edge modification of 2D materials are of great scientific and technological interest
because such edge modification are expected to significantly change the properties of
2D structures, especially 2D nanostructures because of a large edge-to-surface ratio, in analogy to surface
modification of 3D nanostructures with a large surface-to-volume ratio. One well studied example is graphene
nanoribbons (GNRs) \cite{1,2,3,4}. However, much less attention has been paid to edge modification of 2D
topological insulators (TIs). This is because the topological nature of TI edge states, with a origin from
bulk band topology, is well-known to be robust, insensitive to nonmagnetic chemical and structural edge
modification \cite{5,6}. Especially, some fundamental topology-defined transport properties of TI edge state,
such as its quantized edge conductance and spin-momentum locking relation, cannot be changed. On the other hand,
other characteristic transport properties of TI edge state, such as carrier mobility, the number of quantum conductance
and the spin texture, can all in principle be changed by edge modification, but modification of transport properties of
TI edge state has been rarely demonstrated.

Ultrathin Bi(111) films are theoretically predicted to be a 2D TI material \cite{20,21,22,23,zhou}, which have been
experimentally synthesized and characterized recently \cite{27,28,29,30}. In this Letter, we demonstrate that
the transport properties of TI edge states in zigzag Bi(111) nanoribbon (ZBNR) can be tuned by chemical edge
modification via H adsorption. Most remarkably, the Fermi velocity
of Dirac edge state is increased by as much as one order of magnitude, when the Dirac point is moved from
the Brillouin zone boundary to Brillouin zone center. Correspondingly, the real-space distribution
of Dirac states is found to be much more delocalized. In addition, the spin texture of the Dirac states is also modified,
with the spin orientation switching from predominantly in-plane to out-of-plane alignment around the Dirac point.
Through a systematic analysis based on model first-principles and tight-binding (TB) calculations, the physical mechanism
underlying these intriguing phenomena is revealed to be the atomic orbital filtering effect by edge H atoms.
The edge H removes part of x- and z-components of
$p$ orbits of the edge Bi atoms, leading to a change of edge boundary potential that reshapes the topological
edge states. The change of spin texture is further analyzed by an effective Hamiltonian.

The first-principles calculations of ZBNR, containing eighty Bi atoms, including spin-orbit coupling and without/with H
adsorption are carried out in the framework of generalized gradient approximation with Perdew-Burke-Ernzerhof
functional using the VASP package \cite{31}. The supercells have a vacuum layer more than 15 {\AA} thick
to ensure decoupling between neighbouring ZBNRs. All self-consistent calculations are performed with a
plane-wave cutoff of 400 eV on an $1\times11\times1$ Monkhorst-pack k-point mesh. For structural
relaxation, all the atoms are allowed to relax until atomic forces are smaller than 0.01 eV/{\AA}.
The TB calculations of ZBNR are done using the Wannier90 package \cite{32}. First, the TB Hamiltonian of
one unit cell, containing two Bi atoms and twelve maximally localized
Wannier functions (MLWFs) of $p$ orbits, is fitted to the first-principles calculations. Using this unit-cell
TB Hamiltonian, we further constructed a supercell TB Hamiltonian of ZBNR, containing eighty Bi atoms.
To model H adsorption effect, $p$ orbits of edge Bi atoms are selectively removed in the TB
calculations.

Figures 1(a) and 1(b) show the first-principles band structures of ZBNR without and with edge H adsorption, respectively,
illustrating the remarkably different topological edge states under different chemical edge
environments. Without H adsorption [Fig. 1(a)], there are two extended gapless edge states
inside the bulk band gap. The two edge states, connecting the valence and conduction bulk
bands (shaded yellow regions) and forming a 1D Dirac state at the Brillouin zone boundary, are characterized
by an odd number of crossings over the Fermi level (from $k_y=0$ to $k_y=1$). Tuning the Fermi level
within the band gap, the number of crossings can be either three or one, indicating the topological
nature of ZBNR. Due to the inversion symmetry, each edge state is degenerated for the left and right
edges. The real-space distribution of these degenerated edge states at four chosen k-points
[as marked in Fig. 1(a)] are shown in Fig. 1(c). We see that the width of localized edge
state is k-point dependent and its maximum width is $\sim$ 2nm, which is consistent
with previous scanning tunneling microscopy (STM) measurement \cite{28}.

However, with H adsorption at the edge of ZBNR, the topological edge states are dramatically
modified in three important ways, as shown in Figs. 1(b) and 1(d). First, the most significant finding
is that the Fermi velocity ($\nu_F$) of Dirac states, as obviously reflected by the band slope, is increased by
as much as one order of magnitude, changing from $1.1\times10^5$ m/s to $0.9\times10^6$ m/s. This high Fermi
velocity is comparable to the largest $\nu_F$ ($3\times10^6$ m/s) obtained
from suspended graphene \cite{33}. The Fermi velocity is one of the key parameters in the study of
Dirac materials, as it bears various fundamental information, such as electron mobility,
electron-electron interaction \cite{33} and effective fine structure constant \cite{PRL}.
Several different routes have been proposed to engineer $\nu_F$ in graphene \cite{34}, but few
have been reported in TIs. Here, we demonstrate that chemical edge modification presents
a simple and powerful way to engineer $\nu_F$ in 2D TIs.

Second, the original extended gapless edge states become
localized at Brillouin zone center in the $k$-space, similar to the edge states in HgTe/CdTe quantum
well \cite{21}. Consequently, there is only one crossing point between the edge state and  Fermi level
within the energy window of bulk band gap. This modification can have useful implications in spintronic devices.
With one crossing, electron backscattering is in principle completely forbidden. Having three crossing points
the edge state and  Fermi level, electron backscattering can no longer be 100\% prohibited in a scattering
process between two states with same energy but different momenta. This is because the spin directions of those
two states are not completely inverse with each other \cite{28}.

\begin{figure}[htpb]
\begin{center}
\epsfig{figure=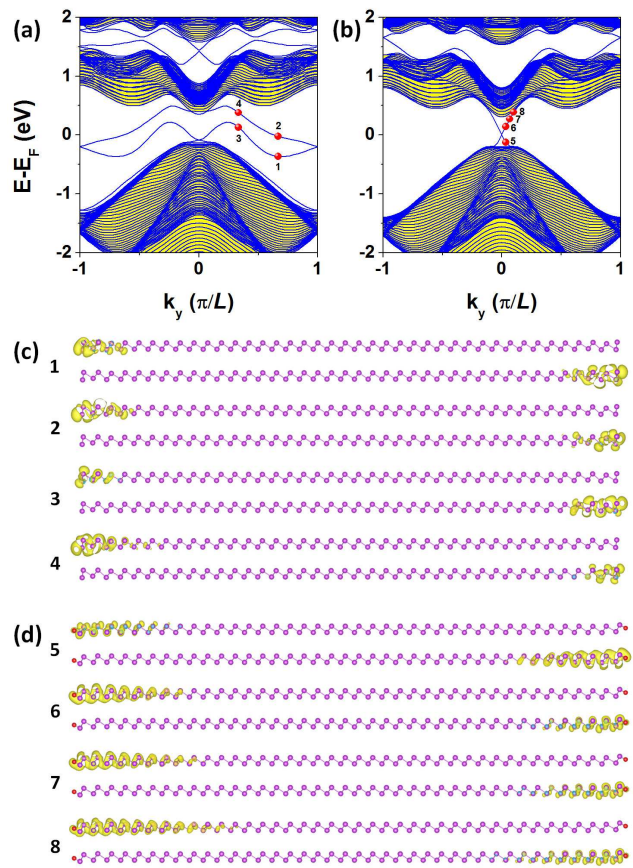,width=8.3cm}
\caption{(a) and (b) First-principles band structures of ZBNR without and with edge H adsorption, respectively.
The shaded yellow regions are bulk states with a band gap; the solid blue lines inside the
band gap are topological edge states. (c) and (d) Side view of real-space charge density
distributions of the edge states at different k-points, as marked in (a) and (b), respectively.}
\end{center}
\vspace{-0.2in} \label{fig:fig-1}
\end{figure}

Third, the real-space distribution of the localized edge states become much wider than those
without H adsorption, as seen by comparing Fig. 1(d) with Fig. 1(c). As k-point moves away from the Dirac point,
the localized edge states become more and more delocalized, whose width of distribution increases to a maximum of $\sim$ 4.4 nm. This
observation is consistent with previous theory of penetration depth of edge states, which is
inverse to k-space distribution of the edge states \cite{21}. A localized state in momentum space will
have an extended distribution in real space.

\begin{figure*}[htpb]
\begin{center}
\epsfig{figure=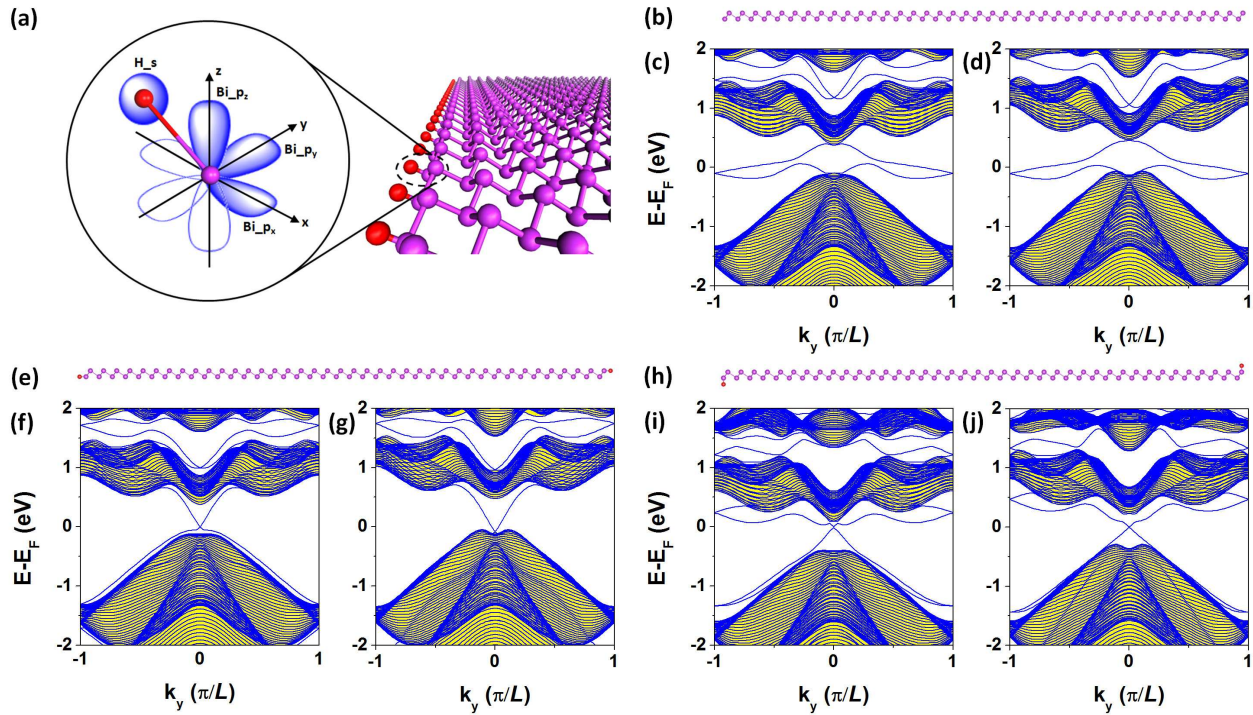,width=16.6cm}
\caption{(a) Illustration of the orbit and bonding at the edge of ZBNR with H adsorption. One s orbit for
each H atom and three $p$ orbits for each Bi atom are shown. (b), (e) and (h) are the atomic structure of ZBNR
without H adsorption, with H adsorption along x-direction and with H adsorption along z-direction, respectively.
The atomic structures are not relaxed. (c), (f) and (i) are the first-principles band structures
of (b), (e) and (h), respectively. (d), (g) and (j) are the TB band structures of (b), (e) and (h),
respectively.}
\end{center} \vspace{-0.2in} \label{fig:fig-2}
\end{figure*}

The results shown in Fig. 1 demonstrate the feasibility of tuning some important characteristic transport properties of 2D TI edge states by
chemical edge modification. Next, we perform an orbital analysis to reveal the physical mechanism
underlying this phenomenon. There are one $s$ orbit for each H atom and three $p$ orbits for
each Bi atom, as schematically shown in Fig. 2(a). At the ribbon edge, each Bi atom is bonded with
a H atom. Structural optimization confirms that the H-Bi bond is within the z-x plane; \textit{i.e.} it is
perpendicular to the $p_y$, but not perpendicular to the $p_x$ or $p_z$ orbit of Bi atom. Considering
the orbital symmetry, it is easy to see that the hopping between $s$ orbit of H and $p_y$ orbit of
Bi is zero, but the hopping between s orbit of H and $p_x$ (or $p_z$) orbit of Bi is not zero.
Therefore, the effect of H atoms is to partially saturate $p_x$ and $p_z$ orbits of edge
Bi atoms.

In order to support such hypothesis, we have further performed several first-principles
and TB model calculations. First, we construct a TB Hamiltonian of ZBNR using the ideally
bulk-terminated edge without edge structural relaxation [Fig. 2(b)]. The band structures obtained
from the first-principles and TB model calculations are shown in Figs. 2(c) and 2(d), respectively.
We see good agreement between the two methods for both bulk and edge states,
validating the effective TB Hamiltonian. Next, we artificially fix the adsorbed
H atoms along either x- or z-direction without structural relaxation, while maintaining
the inversion symmetry of the ribbon, as shown in Figs. 2(e) and 2(h), respectively.
This special setup allows us to selectively saturate either ${p_x}$ or ${p_z}$ orbit of
the edge Bi atoms. Thus, we can distinguish the effect of those two orbits from each other.
The H-Bi bond length is set to 1.82 {\AA}, as obtained from the relaxed structure in Fig. 1(d).

The first-principles band structures for the above two different H-adsorbed model
configurations are shown in Figs. 2(f) and 2(i), respectively. The most significant finding
is that a new Dirac state appears at Brillouin zone center in both cases. In the x-direction H-adsorbed
configuration when the $p_x$ orbit of edge Bi atoms is removed [Fig. 2(f)], the upper branch of the new
Dirac state spans the whole energy window of the bulk band gap, while the lower branch of the new
Dirac state almost merges into the bulk valence bands. However, in the z-direction H-adsorbed configuration when the $p_z$
orbit of edge Bi atoms is removed [Fig. 2(i)], the lower branch of the new Dirac state spans the
most energy window of bulk band gap. Both upper and lower branches of the new Dirac state have an
extended dispersion, resulting in two additional Dirac states at Brillouin zone boundary.

To more clearly see the H adsorption effect in above two configurations, we also selectively removed
the $p_x$ or $p_z$ orbit of edge Bi atoms in the TB Hamiltonian. This can be done by adding a
large on-site energy on these selected orbits, similar to a method used previously to study the
gapless edge state in GNR \cite{yao}. The corresponding TB band structures are shown in
Figs. 2(g) and 2(j) for removing $p_x$ and $p_z$ orbit, respectively, which are consistent with
the first-principles results of Figs. 2(f) and 2(i). The comparison between the first-principles
and TB calculations indicates that the H atoms act as an orbital filter \cite{zhou}, which selectively removes
the orbit of edge Bi atoms and reshape the topological edge states. It is the combined
effect of partial removal of both $p_x$ and $p_z$ orbits of edge Bi atoms that is responsible
for generating the new gapless Dirac edge states in Fig. 1(b) by H edge adsorption.

It is interesting to note that if both $p_x$ and $p_z$ orbits of edge Bi atoms were completely removed simultaneously,
e.g., by adsorbing two H atoms on the edge Bi atoms, the Dirac state would remain located at Brillouin zone boundary, as shown in Fig. S1(a) \cite{35}.
Also, the effect of removing $p_y$ orbit of edge Bi atoms has been studied. It would reshape the edge states and induce two Dirac
states at Brillouin zone boundary, as shown in Fig. S1(b) \cite{35}. However, practically it is hard to find a way to just remove the $p_y$ orbit.
Furthermore, if there is no chemical adsorption, structural relaxation alone will not change the position of Dirac point, and
hence without significantly changing the shape of edge state. This can be seen by
comparing the first-principles band structures of ZBNR with [Fig. 1(a)] and without [Fig. 2(c)]
structural relaxation.

Next, we explain where the new gapless Dirac edge state at Brillouin zone center comes from. The TB
model provides us with an easy method to analyze the evolution of topological edge
states upon edge modification. We can continuously remove $p_x$ and $p_z$ orbits of edge Bi atoms by increasing
their on-site energies gradually in the TB Hamiltonian. As shown in Figs. 3(a)-(c), by gradually
increasing the on-site energy of $p_x$ orbit, the edge states are continuously reshaped. The upper branch of
the old Dirac state merges into the conduction bands, and the lower branch of the old Dirac state
becomes the upper branch of the new Dirac state, locating the Dirac point at Brillouin zone center. However, the case for
$p_z$ orbit is a little different, as shown in Figs. 3(d)-(f). By gradually increasing the on-site
energy of $p_z$ orbit, the edge states are continuously moved upward. Part of the old Dirac
state merges into the conduction bands, and the new Dirac state is pulled up from the valence bands.
The bulk states are not changed in both cases. From this model calculation, we found that depending
on the edge potential, the topological edge states can be either regenerated from or merged into the bulk states,
always being connected to bulk states.

Lastly, we show that the spin textures of topological edge states can also be modified by chemical edge modification.
The zoom-in first-principles band structures and spin textures around the Dirac states without and with H
adsorption are shown in Figs. 4(a) and 4(b), respectively. We see that the spins are within the z-x plane
and perpendicular to the momentum direction (y-direction), showing the helical nature and spin-momentum locking
property. Without H adsorption [Fig. 4(a)], the in-plane (x-direction) spin component is dominant
near the Dirac point. When k-point goes away from the Dirac point, the out-of-plane (z-direction) spin
component becomes larger. With H adsorption [Fig. 4(b)], the out-plane (z-direction) spin component is
dominant near the Dirac point. When k-point goes away from the Dirac point, a small in-plane (x-direction)
spin component starts to appear. Thus, the spin direction is wrapping along the Dirac state with different momenta
in the two cases. Such a spin texture of the 1D Dirac states can be described by the following low-energy
effective Hamiltonian,
\begin{equation}
H=s\hbar\nu_Fk_y[\lambda(k_y)\sigma_x+\sqrt{1-\lambda^2(k_y)}\sigma_z].
\end{equation}
where $s=\pm1$ denotes right/left edge, $\nu_F$ is the Fermi velocity and $\sigma_{x/z}$ is
the pauli matrix. $\lambda\in[-1,1]$; it is a wrapping function of momentum $k_y$. If
$\lambda=0$, the spin only has z-component. If $\lambda=\pm1$, the spin only has x-component.
In other cases, the spin can have both x- and z-components, and spin direction
is tuned by the value of $\lambda$.

\begin{figure}[htpb]
\begin{center}
\epsfig{figure=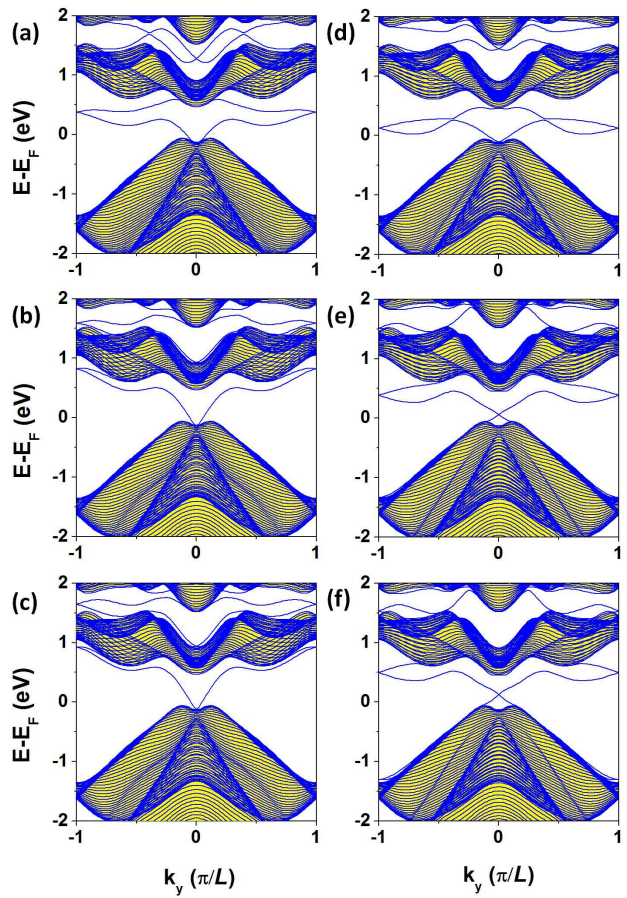,width=8.3cm} \caption{(a)-(c) Evolution of the topological
edge states by gradually removing $p_x$ orbit of the edge Bi atoms. (d)-(f) Evolution
of the topological edge states by gradually removing $p_z$ orbit of the edge Bi atoms.
From (a)-(c) and (d)-(f), the on-site energy for $p_x$ and $p_z$ orbits
are set at 1, 3 and 5 eV, respectively.}
\end{center} \vspace{-0.2in} \label{fig:fig-3}
\end{figure}

\begin{figure}[htpb]
\begin{center}
\epsfig{figure=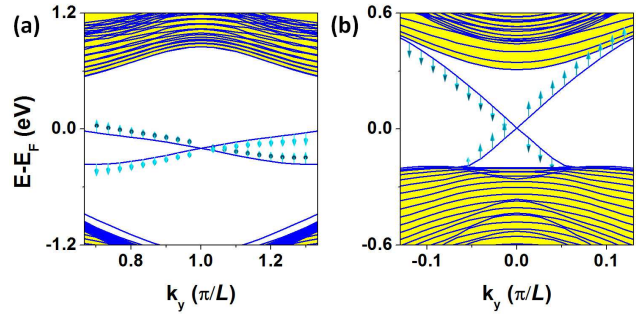,width=8.3cm} \caption{(a) and (b) Zoom-in band structures around
the Dirac point for Fig. 1(a) and Fig. 1(b), respectively. The arrows, perpendicular to momentum,
denote the spin textures of edge states.  Only the left edge-state spin textures are plotted
for each band, which are inverse to the right edge-state spin textures.}
\end{center} \vspace{-0.2in} \label{fig.fig-4}
\end{figure}

In conclusion, although the topology of TI edge states is robust against
structural and chemical edge modification, we show that some important characteristic transport properties
of TI edge states can still be significantly modified, as illustrated by H edge adsorption
in ZBNR. Most remarkably, the Fermi velocity can be increased by as
much as one order of magnitude, becoming comparable to the largest value found in graphene.
Also, the spin textures of TI edge states are modified. Our findings are scientifically interesting
for a better understanding of basic transport properties of topological edge states in relation to edge boundary conditions,
in addition to bulk topology. They also have significant practical implications, because those modified
transport properties, such as Fermi velocity and spin texture, are important parameters in TI-based spintronic devices. The approach of
engineering topological edge states by chemical edge modification is general, not only applicable to other 2D TIs but can also
be extended to 3D TIs via chemical surface modification.


\end{document}